\documentclass[acus]{JAC2003}

\usepackage{graphicx}

\begin{document}

\title{BEAM PROFILE MEASUREMENTS AND SIMULATIONS OF THE PETRA LASER-WIRE}

\author{J. Carter, I. Agapov, G. A. Blair, G. Boorman, C. Driouichi, F. Poirier \\
M. T. Price (Royal Holloway University of London, Surrey), T. Kamps (BESSY GmbH, Berlin),\\
  K. Balewski, H. Lewin, S. Schreiber, K. Wittenburg (DESY, Hamburg),\\
   N. Delerue, D. F Howell (University of Oxford, Oxford), S. T. Boogert, S. Malton (UCL, London)
   \thanks{This work is supported by the Commission of European Communities under the 6th Framework Programme "Structuring the European Research Area", contract number RIDS-011899. We also acknowledge support from the Royal Society}}

\maketitle

\begin{abstract}
The Laser-wire will be an essential diagnostic tool at the International Linear Collider. It uses a finely focussed laser beam to measure the transverse profile of electron bunches by detecting the Compton-scattered photons (or degraded electrons) downstream of where the laser beam intersects the electron beam. Such a system has been installed at the PETRA storage ring at DESY, which uses a piezo-driven mirror to scan the laser-light across the electron beam. Latest results of experimental data taking are presented and compared to detailed simulations using the Geant4 based program BDSIM.
\end{abstract}

\section{INTRODUCTION}

The International Linear Collider (ILC) will be a TeV-scale lepton collider that will require non-invasive beam size monitors with micron and sub-micron resolution for beam phase space optimisation \cite{Ross:2003pa}. Laser-wire monitors operate by focussing a laser to a small spot size that can be scanned across the electron beam, producing Compton-scattered photons (and degraded electrons). These photons can then be detected further downstream using the total energy observed as a function of the laser spot position to infer the transverse profile of the electron bunch. The Laser-wire system installed in the PETRA ring is part of an ongoing effort in the R\&D of producing a feasible non-invasive beam size diagnostic tool.

\section{EXPERIMENTAL SETUP}

The PETRA accelerator was chosen for the installation of the Laser-wire experiment because it is capable of producing bunch patterns similar to the ILC. Laser-wire tests are run using a $7~\mbox{GeV}$ positron beam with a single bunch with a charge of typically $7.7~\mbox{nC}$. From the optics lattice the average beam size is $\sigma_x=268~\mbox{$\mu$m}$ for the horizontal and $\sigma_y=68~\mbox{$\mu$m}$ for the vertical dimension.

\begin{figure}[htb]
\centering
\includegraphics*[width=65mm]{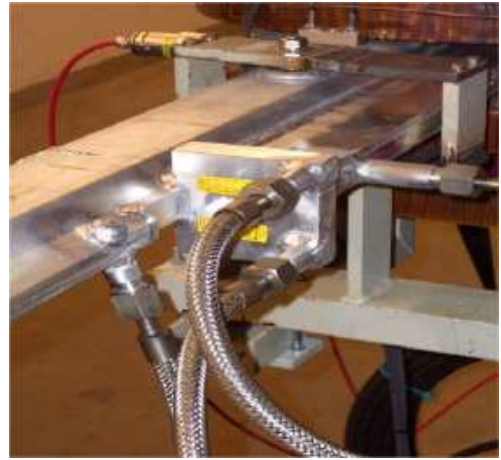}
\caption{New exit window for Compton photons}
\label{fig-PETRA_window}
\end{figure}

Preliminary simulations showed that the Compton-scattered photons loose the majority of their energy in the material of the dipole magnet's beampipe due to hitting the wall with a shallow angle, resulting in an effective length of $100~\mbox{cm}$ of Aluminium. An exit window was therefore designed and installed (by DESY) to allow these photons to reach the detector with little deterioration (see Fig.~\ref{fig-PETRA_window}).

\subsection{Laser Beam}

The laser pulses are created in a Q-switched
Nd:YAG laser operating at $532~\mbox{nm}$.
The pulses are then transported via a matched
Gaussian relay made up of two lenses over a distance of $20~\mbox{m}$ from
the laser hut via an access pipe
into the tunnel housing the accelerator.
The laser beam is then reflected off the scanning
mirror before it reaches a focusing lens
with $f = 117~\mbox{mm}$ back-focal length.
The scanner is a piezo-driven platform with
an attached high-reflectivity mirror which has a maximum scan range of $\pm 2.5~\mbox{mrad}$.
The peak power at the laser exit was measured
to be $3.63~\mbox{MW}$. At the IP the peak
power is reduced to $1.46~\mbox{MW}$ as higher order modes carry some fraction of the beam
power but these are focussed out of beam transport,
which is only matched for the fundamental mode.
The longitudinal
profile was measured using a streak camera with
$5~\mbox{ps}$ time resolution. The data revealed a
pulse length of $\Delta t = 12.5~\mbox{ns}$ FWHM with
a sub-structure of roughly $70~\mbox{ps}$ peak-to-peak
and $70~\mbox{ps}$ peak width at full contrast due to mode-beating. This causes the Compton signal
amplitude to vary between zero and full signal for
different laser shots. In order to reduce the data taking time
the current laser will be replaced with an injection
seeded system enabling faster data taking.

\subsection{Compton Calorimeter}

The Laser-wire set up makes use of a calorimeter composed of 9 Lead Tungstate ($PbWO_4$) crystals arranged in a $3 \times 3$ matrix fixed with optical grease to a square faced photomultiplier. The individual crystals have dimensions of $18 \times 18 \times 150~\mbox{mm}$. The complete detector set up was tested with a testbeam from the DESY II accelerator using electrons from $450~\mbox{MeV}$ to $6~\mbox{GeV}$. Energy resolution was shown to be better than $6\%$ for individual crystals and $10\%$ for the overall set up. Simulations show that for the $3 \times 3$ matrix, $95\%$ of the total energy deposit is collected for an incoming Compton-scattered photon with $300~\mbox{MeV}$ energy \cite{Blair:2002rn}.

\subsection{Data Acquisition}
The Laser-wire DAQ system has two main components: the hardware trigger which synchronises the laser and DAQ components to the electron (positron) bunch, and the software which controls the acquisition and collation of data from each sub-component of the system.

The hardware trigger operates with two inputs from the PETRA Integrated Timing system (PIT) and produces the necessary signals to fire the laser. The trigger card also produces a signal to trigger the CCD cameras and a signal to start the software data acquisition.
When the signal from the trigger card is received a counter which runs for approximately $420~\mbox{$\mu$s}$ is started.  After this time a signal is sent to the integrator card, lasting around $50~\mbox{$\mu$s}$, to integrate the output from the calorimeter.  The integrated signal is read by an ADC channel.

The DAQ software also produces a programmable signal, up to a peak of $10~\mbox{V}$, which is amplified by a factor of 10 and this is used to drive the piezo-electric scanner.  A scaled version of the scanner amplifier output is read by an ADC channel. The other sub-components of the DAQ system: the BPM monitor, the PETRA data monitor and the CCD cameras are also read out.  Communication with each component is performed by a messaging system using TCP/IP.

\section{DATA ANALYSIS}

\subsection{Laser Beam Size}
In order to determine the transverse size of the electron beam, it is necessary
to know the properties of the laser that is being used to scan. Particular attention is paid to
the spot size at the laser waist, $\sigma_0$, and the
Rayleigh range, $z_R$, (the distance from the waist at which the beam size
 $\sigma = \sqrt{2} \sigma_0$). These properties are related by Eq.~\ref{equ-laserprofile}:
\begin{equation}\label{equ-laserprofile}
 \sigma = \sigma_0 \sqrt{1+\left(\frac{z}{z_R}\right)^2}
\end{equation}
where $z_R = \frac{4 \pi \sigma_0^2}{M^2 \lambda}$.\\

The laser is focused using the same final focus lens as described previously. A CMOS camera is placed
on a track rail so that it can be moved through the focal plane parallel to
the beam direction.
Due to the high power of the laser, the beam was first passed through a
$99.9~\mbox{\%}$ reflective mirror, and then through a variable amount of
neutral density filter in order to prevent saturation and damage to the
camera pixels. The camera was moved along the track rail to a number of
positions, and 100 images were taken in each location.

\begin{figure}[htb]
\centering
\includegraphics*[width=65mm]{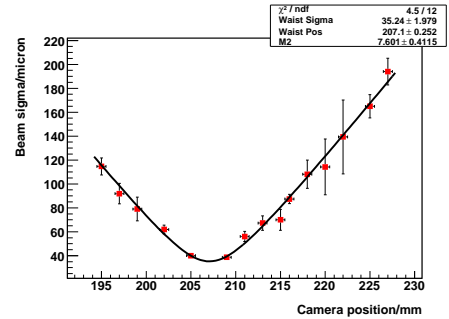}
\caption{Variation in transverse beam size of the laser around the focus at the IP.}
\label{fig-laserwaist}
\end{figure}

The images taken by the camera are stored as 8-bit greyscale bitmap files.
The pixel data is projected onto the y-axis, and fitted to a gaussian
on a linear background in the region around the signal peak. The width
at each location is then plotted, and fitted to Eq.~\ref{equ-laserprofile}.
From this we obtain $M^2=7.6 \pm 0.41$, which is within the expected range, and
$\sigma_0=(35\pm2)~\mbox{$\mu$m}$, as shown in Fig.~\ref{fig-laserwaist}.

\subsection{Scan Data}

The laser is scanned across the electron beam by tilting a mirror on a
piezo-electric stack to produce a deflection of $\pm2.5~\mbox{mrad}$.
Focusing through the lens produces a travel range for the focal spot at the IP of
$585~\mbox{$\mu$m}$.
The scanner voltage is applied in a stepped sinusoidal pattern; $10$ triggers
are taken at each of $100$ voltages over a whole $2\pi$. The trigger signal is
taken from the laser trigger card running at $30~\mbox{Hz}$, so a full scan
takes approximately $33~\mbox{s}$.

\begin{figure}[htb]
\centering
\includegraphics*[width=65mm]{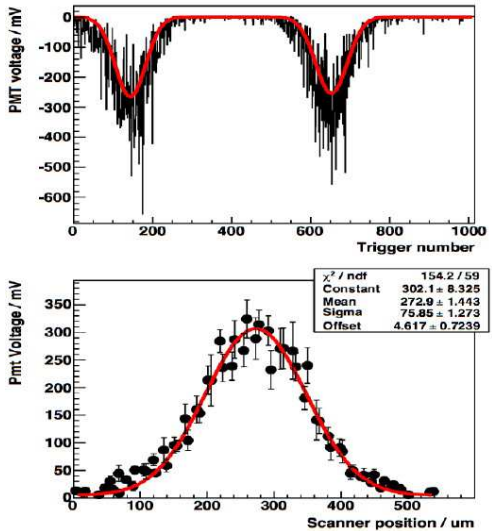}
\caption{a: PMT voltage vs trigger number, fitted to a constant background with two gaussians. b: Mean PMT voltage vs laser focus position with a fit showing $\sigma_m$.}
\label{fig-scanplots}
\end{figure}

The signal from the ADC is expected to display two peaks; one as the laser
crosses the electron beam on a rising voltage to the scanner, and one on
a falling voltage. The trigger number exactly half way between the peaks
should correspond to a turning point in the scanner position. The mean of the background subracted ADC counts
at each voltage is then fitted to a gaussian whose width, $\sigma_m$ is given
by $\sigma_m^2 = \sigma_e^2 + \sigma_0^2$.  Fig.~\ref{fig-scanplots} shows the typical results observed for a single scan and the results of multi-scan shifts are presented in Table~\ref{tbl-eresults}. Note that the large signal variation in Fig.~\ref{fig-scanplots}a is partly due to the sub-structure of each laser pulse and will be removed by a better laser.

\begin{table}[hbt]
\centering
\begin{tabular}{| c | c | r |}
\hline
\textbf{Shift} & \textbf{No. of Scans} & \large{$\sigma_e$} [\mbox{$\mu$m}] \\ \hline
1 & 7 & $62.89\pm2.45$ \\
2 & 7 & $71.67\pm3.28$ \\
3 & 3 & $77.22\pm5.51$ \\ \hline
\end{tabular}
\caption{Data run results for extracted electron beam size, $\sigma_e = \sqrt{\sigma_m^2 - \sigma_0^2}$. The errors are the RMS from several scans}
\label{tbl-eresults}
\end{table}

\section{COMPARISON WITH SIMULATIONS}

The entire PETRA Laser-wire set up has been simulated using BDSIM \cite{Blair:BDSIM}, which is a fast tracking code utilising the Geant4 \cite{Geant4:weblink} physics processes and framework. The simulation is a full model of the accelerator components including beampipe, magnets, and cooling water channel. For each simulated event a Compton scattered photon is generated with an energy based upon the Compton Spectrum predicted for the PETRA Laser-wire parameters. This photon is tracked to the detector whilst fully simulating any interactions with materials such as the beampipe wall. This process is repeated to create an effective single Compton energy distribution and its corresponding distribution at the detector after passing through any matter along the photon path.

\begin{figure}[htb]
\centering
\includegraphics*[width=65mm]{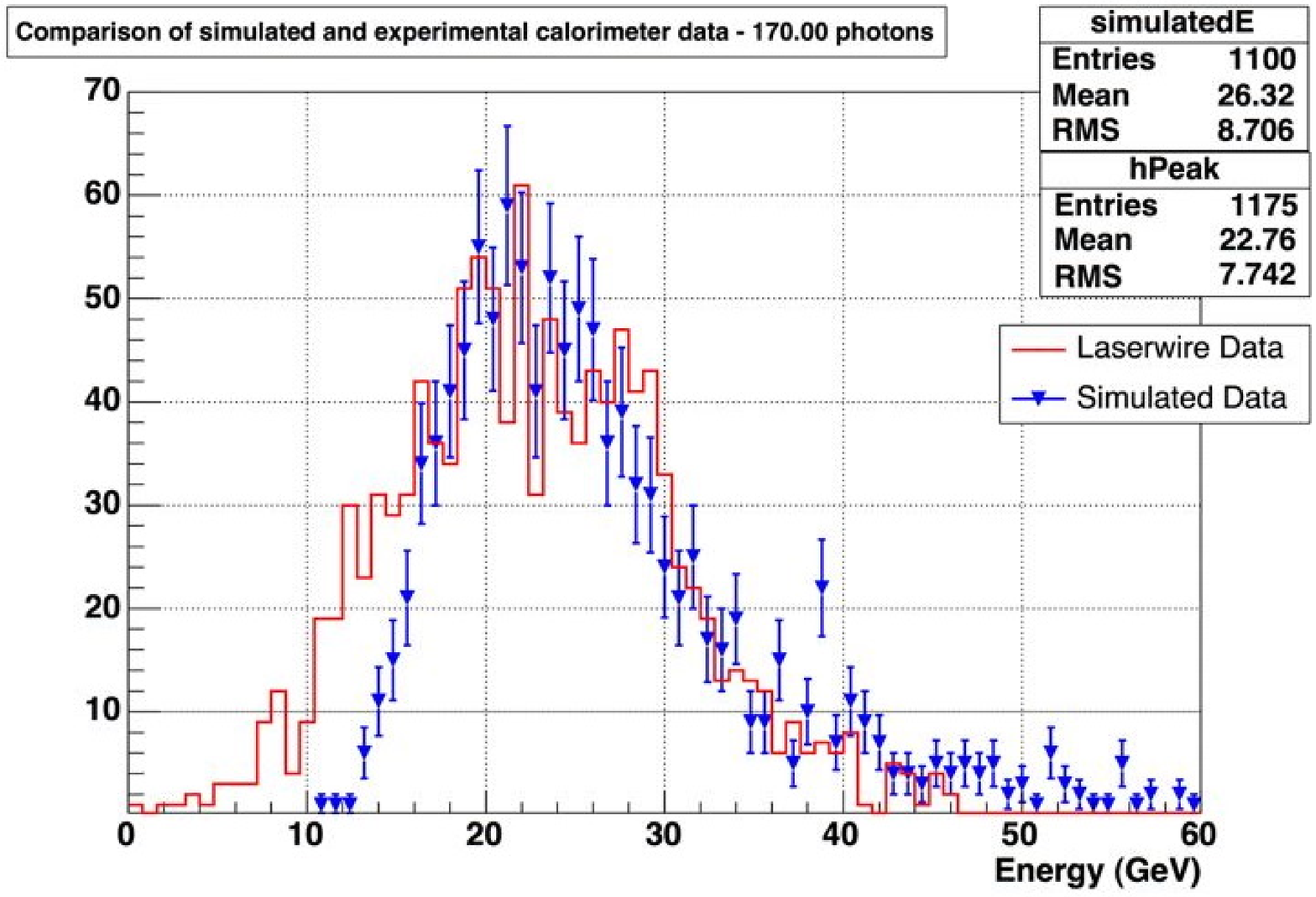}
\caption{Calorimeter energy spectra for data and simulated events.}
\label{fig-sim}
\end{figure}

The single photon distribution in the detector is extrapolated to the $N_{photon}$ spectrum using Poisson statistics whilst also accounting for the energy resolution of the calorimeter and the longitudinal sub-structure of a  typical laser pulse. The simulated spectrum is compared directly to the experimental data (see Fig.~\ref{fig-sim}), where the laser and electron beam were well aligned. The expected number of Compton-scattered photons, $N_{photon}$, per shot with the Laser-wire setup parameters is approximately $170\pm25$ photons, which agrees with the theoretical value.

The experimental data show an energy resolution of 34\% which is dominated by the longitudinal fluctuations in the laser power. The simulation models these fluctuations using relatively old streak camera data as described above and so does not account for degradation in the quality of the laser since then. The calorimeter has also not been calibrated for the range of energy deposits now incident upon it and has been in the PETRA radiation environment for three years. This could explain why the simulation fails to completely model the experimental data in the lower energy region.

\section{OUTLOOK}

The future strategy for the Laser-wire project can be characterised
in the short term to concentrate on non-laser issues like
data acquisition, signal detection, vertical scanning, and implementation
into a linac beamline. This aims at the development of
a standard diagnostic tool to be placed at many locations
along the accelerator beamline.
In the long run R\&D work is planned to develop a laser
system producing pulses matching the ILC micro pulse structure.
Here the target is to have a beamsize monitor with
full flexibilty.
To meet the short-term targets it is planned to purchase
an injection seeded Q-switch laser with second harmonic
generation having excellent longitudinal and transverse
mode quality. A complimentary project concentrating on the achievement of micron-scale laser spot-sizes is underway at the Accelerator Test Facility (ATF) at KEK.

\end{document}